\documentstyle[aps,prl,manuscript]{revtex}

\begin{document}

\title{Improved upper limit on 
Muonium
to Antimuonium Conversion}
\author{
R.~Abela$^1$,
J.~Bagaturia$^2$,
W.~Bertl$^1$,
R.~Engfer$^3$,
B.~Fi\-scher von Weikersthal$^4$,
A.~Gro\ss{}mann$^4$,
V.W.~Hughes$^5$,
K.~Jungmann$^4$,
D.~Kamp\-mann$^6$,
V.~Karpu\-chin$^7$,
I.~Kisel$^7$,
A.~Klaas$^6$,
S.~Korenchen\-ko$^7$,
N.~Kuchinsky$^7$,
A.~Leuschner$^3$,
B.E.~Matt\-hias$^4$,
R.~Menz$^3$,
V.~Meyer$^4$,
D.~Mzavia$^2$,
G.~Otter$^6$,
T.~Prokscha$^4$,
H.S.~Pruys$^3$,
G.~zu~Putlitz$^4$,
W.~Reichart$^3$,
I.~Reinhard$^4$,
D.~Renker$^1$,
T.~Sakhelash\-vil\-li$^2$,
P.V.~Schmidt$^4$,
R.~Seeliger$^6$,
H.K.~Walter$^1$,
L.~Willmann$^4$,
L.~Zhang$^4$
}
\address{
$^1$ Paul Scherrer Institut, CH-5232 Villigen PSI, Switzerland
\newline
$^2$ Tbilisi State University, GUS-380086 Tbilisi, Georgia
\newline 
$^3$ Physik Institut, Universit\"at Z\"urich, CH-8057 Z\"urich, Switzerland
\newline
$^4$ Physikalisches Institut der Universit\"at 
Heidelberg, D-69120 Heidelberg, Germany
\newline
$^5$ Physics Department, Yale University, New Haven Ct., 06520, USA
\newline
$^6$ Physik Institut B, RWTH Aachen, D-52056 Aachen, Germany
\newline
$^7$ Joint Institute of Nuclear Research, RU-141980 Dubna, Russia
\newline
}

\maketitle

\begin{abstract}
{
A new experiment has been set up at the Paul Scherrer Institut
to search for muonium to antimuonium conversion.
No event was found to fulfil the requested signature 
which consists of the coincident detection of both
constituents of the antiatom in its decay.
Assuming an effective (V-A)$\times$(V-A)
type interaction
an improved upper limit 
 is established for the conversion probability 
of ${\rm P_{M\overline{M}}}
\leq 8 \cdot 10^{-9}$ (90\%C.L.), 
which is almost two orders of magnitude lower compared to 
previous results and
provides a sensitive test for theoretical
extensions of the standard model.
}
\end{abstract}



The hydrogen-like muonium atom (M=$\mu^+ e^-$) consists of two
leptons from different generations.
Due to the close confinement of the bound state 
it offers excellent opportunities
to study precisely the fundamental electron-muon interaction
as described in standard theory and
to search sensitively for additional so far unknown
interactions between these two particles.
A spontaneous conversion of muonium into antimuonium (${\rm
\overline{M}} = \mu^- e^+)$~ would violate additive lepton family
number conservation by two units.
In the standard model, which is a very successful description of experimental 
particle physics, this process is not provided
like other decays which are searched for, e.g.
the muon decay modes $\mu^+ \rightarrow e^+\nu_\mu\overline\nu_e$\cite{hwhite},
$\mu \rightarrow e\gamma$\cite{mega}, $\mu \rightarrow eee$\cite{bert85} and 
$\mu \rightarrow e$ conversion\cite{schaaf}. 
However, in the framework of many speculative theories, which try to 
extent the standard model in order to 
explain further some of the
features like parity violation in weak interaction or the particle
mass spectra,
lepton number violation appears to be natural and
muonium to antimuonium conversion is an essential part in several 
of those models (Tab. 1)\cite{herc92,halp82,wong94,moha92,halp93,fuji94}.
The coupling constant ${\rm G_{M\overline{M}}}$ in an effective 
four fermion interaction \cite{fein61} could be as large as
the present experimental limit
of ${\rm G_{M\overline{M}}} \leq 0.16 {\rm G_F}$~ (90\%C.L) 
established at LAMPF in Los Alamos, USA, \cite{matt91} or the bound 
of ${\rm G_{M\overline{M}}} \leq 0.14 {\rm G_F}$~ (90\%C.L)
\cite{gord94} very
recently proposed from
an experiment at the Phasotron in Dubna, Russia, where
${\rm G_{F}}$ is the Fermi coupling constant of the weak interaction.
In particular in the framework of minimal left-right symmetric theory a lower
bound has been predicted with the assumption of a muon neutrino mass
$m_{\mu_\nu}$ larger than 35 keV/c$^2$.

At the Paul Scherrer Institut (PSI) in Villigen,
Switzerland, a new experiment has been
set up (Fig. \ref{mmbarsetup}), which
utilizes the powerful signature for a conversion developed in the recent
LAMPF experiment \cite{matt91}. 
It requires the coincident identification of both
constituents of the antiatom, $\mu^-$ and $e^+$,
in its decay.
Since a possible conversion is suppressed in matter mainly 
due to the removal of degeneracy
of between muonium and antimuonium \cite{fein61,morg70},
a sensitive experiment 
the needs the atoms to be in vacuum.
This experiment utilizes
the most efficient method known to date to provide
muonium in vacuum: Positive muons from a continuous beam of momentum
p=21 MeV/c and momentum bite $\Delta$ p/p = 6 \% at a typical 
beam rate of $8*10^5$ /s are stopped in a silicon
dioxide ($SiO_2$) powder target of thickness 8 mg/cm$^2$. 
About 60 \%
of the muons form muonium atoms \cite{kief82}, some of which leave the target 
through its
surface into the surrounding vacuum with thermal
energies, corresponding to a velocity of 
7.4(1) mm/$\mu$s \cite{wood88}.

Electrons from the decay $\mu^- \rightarrow e^- +\overline{\nu_e} + \nu_\mu$
of an antimuonium atom
have an energy spectrum ranging up to 53 MeV. They
can be observed in a
magnetic spectrometer
consisting of five
cylindrical proportional chambers surrounded by a
hodoscope of 64 plastic scintillator
stripes.
Three of the wire chambers are equipped with two
planes of helical cathode stripes
allowing to measure both, angular and axial, coordinates \cite{bert85}.
A solenoidal magnet provides an axial magnetic field of
0.1T for measuring particle momenta and identification of their 
sign of charge.
After the decay of the antiatoms the positrons from
their atomic shells are left behind with an average
kinetic energies of $R_{\mu}$ = 13.5 eV \cite{chat92}.
These positrons can be 
accelerated parallel to the magnetic flux lines up
to typically 8 kV in a two stage electrostatic device and
guided in a magnetic field of $B=0.1$ T
onto a position sensitive microchannel plate detector
(MCP) with resistive anode readout \cite{MCP96}. 
The transport system consists of two straight sections of 1.5 m in lenght 
with a 90 $^\circ$ bend of radius 35 cm between them (Fig. \ref{mmbarsetup})
and it is momentum selective with a transmission maximum for 8 keV positrons.
An electric field orthogonal to the
magnetic field applied in the first section 
displaces charged particles by a distance depending on their velocity
and removes slow ions or muons.
Particles of longitudinal momenta exceeding
750keV/c cannot follow adiabatically the magnetic
flux lines in the bend region. They gain transverse momenta
above 135keV/c resulting in a
gyration radii larger than 4.5 mm
and they are stopped in a
collimator of length 40 cm made from
copper foils of  thickness 1 mm 
with a spacing of 10 mm.
The magnetic field gradient causes in addition a drift
orthogonal to both the field and its gradient 
for charged particles
proportional to their momentum.

The positron track can be retraced 
from the position measured on the MCP detector 
to the target region
with an accuracy of 1mm.
The decay position of the atom can be determined from the
vertex with the track of the high energetic particle
in the magnetic spectrometer.
In addition, the detection of at least one of
the 511keV photons from
positron annihilation on the MCP in a 12-fold segmented 
pure CsI crystal
detector surrounding the MCP is required as a part of the signature.
The crystal detector was calibrated
using positrons from a $^{22}Na$ source. 
The acceptance for at least one of the annihilation photons has been 
measured  to be 79(4) \%.
The time resolution of the device is 5.4(3) ns (FWHM).

The new experiment was designed to provide as high as 
possible symmetry 
in the detection of
muonium and antimuonium decays in order to minimize the influence
of systematic uncertainties arising from corrections for 
efficiencies and acceptances of
various detector components.
In the course of data acquisition the polarity of all electric and magnetic
fields were
reversed regularly every four hours for a duration of
20 minutes to monitor the muonium
production and for checking the calibration 
of the detector components and the transport system parameters.
In this case energetic
positrons and atomic electrons from muonium decays were detected.
For a time of flight measurement the time interval between the arrival of a
muon signaled by the beam counter and the detection of its
decay positron in the hodoscope 
was recorded at low beam rates.
On average every second day the $SiO_2$ targets were replaced to compensate for
an observed decrease of the muonium yield.
The time of flight for the electrons from the atomic shell
was determined in to be 
75.8(1)ns by observing muonium atom decays at 8 kV acceleration voltage.
The signal
width of 5.7(1)ns (FWHM) allows to apply a 20ns narrow
coincidence time window in the event signature.

The evaluation of the muonium production data is based an a model established
in independent dedicated experiments \cite{wood88}, which assumes that the
atoms are produced inside of the $SiO_2$ powder at the positions given by the 
stopping distribution of the muons. A one dimensional
diffusion process describes the escape of the muonium atoms 
into vacuum where their 
velocities follow a 
Maxwell-Boltzmann distribution.
The number of atoms in the fiducial volume has been extracted from
two different projections.
Firstly, the distribution of time intervals between a beam counter signal
from the incomming muon and the detection of the atomic electron on the MCP
yields the fraction of muonium in vacuum (Fig. 2a). 
Secondly, this quantity has been derived 
by a two-dimensional maximum likelyhood fit to the distribution
of the atoms decay positions downstream of the target
and their respective time of flight (Fig 2b).
Both methods agree within their error margins and yield
an average fraction of observed muonium atoms
of $3.1(2)\cdot 10^{-3}$ per
incoming muon. This figure includes
all detection efficiencies, i.e. the solid angle of the
magnetic spectrometer ($0.71 \times 4\pi$),
the track reconstruction efficiency (0.85), the MCP detection
efficiency (0.16(2)) and the geometric acceptance of the
transport system for electrons (0.80).
The average fraction of muonium atoms in vacuum per incoming muon amounts to
5.7(8)\%, in good agreement with results from the earlier dedicated 
studies of muonium production\cite{wood88}.

The effective measurement time for the search for antimuonium was 210 hours
during which  $1.4(1)\cdot10^{9}$  muonium atoms were in the fiducial
volume of diameter 9 cm and length 10 cm.
Data were recorded at acceleration voltages between 2 and 10kV to allow for
systematic studies of the transport system. 
No decay of an antimuonium atom was found. There is no
entry in a 20 ns time window around the expected time of flight in the
corresponding distribution (Fig. \ref{mbarresult}).
The apparent structure around $t_{TOF}-t_{expected}$ =-50 ns
corresponds to
the allowed rare
decay mode $\mu^+ \rightarrow e^+e^+e^- \nu_e \overline{\nu_{\mu}}$ 
in which one
of the positrons is released with low kinetic energy, while the electron is
detected in the magnetic spectrometer and partly to Bhabha scattering.
Due to their significantly higher
initial momenta positrons from these processes arrive
at earlier times at the MCP and can be clearly
distinguished from possible antimuonium decays. In addition, for those events
 no decay vertex
can be reconstructed inside of  the fiducial volume,
because of the strong deflection of
high energetic positrons in the bend of the transport system.
The background due to these processes to the search for antimuonium has been 
estimated from Monte Carlo simulations
to be below $10^{-14}$ per incoming $\mu^+$.

Taking
into account the detection efficiency for the annihilation photon in the CsI
detector and 
corrections for the finite fiducial
volume and finite observation time (0.74(2)),
an upper limit can be set on the conversion
probability in a 0.1 T magnetic field
of ${\rm P_{M\overline{M}} (0.1~T) \leq 2.8\cdot 10^{-9}}$
( $90\%$ C.L.).
The conversion is suppressed in external magnetic fields due the removal
of the
degeneracy of the energy levels in muonium and antimuonium
atoms and has been calculated recently for different types of interactions
\cite{wong95,hori95}. For ${\rm (V\pm A)\times(V\pm A)}$ 
interactions,
the conversion probability at a magnetic field of 0.1T
is suppressed to 35\% of the zero field value, while  
for a (V-A)$\times$(V+A) interaction it is reduced to only 77.6\%. In
case of a future observation of muonium-antimuonium conversion the
coupling type could be determined by a measurement of 
the ratio of conversion probabilities at different magnetic field strenghts.
For an effective ${\rm (V\pm A)\times(V\pm A)}$ interaction,
where the zero field probability $\rm P_{M\overline{M}} (0T)$ is related to
the coupling constant ${\rm G_{M\overline{M}}}$ through
\begin{eqnarray*} 
{\rm G_{M\overline{M}}} =\rm G_F \cdot
\sqrt{\frac{\rm P_{M\overline{M}} (0T)}{2.56\cdot10^{-5}}} \;\;\; ,
\end{eqnarray*}
we have
${\rm P_{M\overline{M}}({0T}) \leq 8 \cdot 10^{-9}}$ (90\% C.L.)
and an
upper limit of ${\rm G_{M\overline{M}}} \leq  1.8 \cdot 10^{-2}~\rm G _{\rm F}$
(90\%C.L.).
In GUT models 
where a muonium-antimuonium conversion could be mediated via the 
exchange of a dileptonic gauge boson $X^{\pm \pm}$
a tight new mass limit of
${ M_{X^{\pm\pm}}/g_{3l}} > 1.1$~TeV/c$^2$ (90\%C.L.)
can be extracted where $g_{3l}$ depends on the particular 
symmetry and is of order unity. With 
the limit established in this experiment 
models with dilepton exchange \cite{fuji94} as well as models with heavy
leptons and radiative generation of leptons masses
appear less attractive\cite{wong94}.

With improved detectors for 8 keV positrons which are 
now available with four times enhanced efficiency \cite{MCP96}
and by utilizing the 
beam line $\pi$E5 at PSI with 10 times higher muon fluxes compared to
the $\pi$E3 area used for this experiment we expect 
another significant 
improvement in the sensitivity on the conversion probability in the 
future which could provide a more stringent test
for speculative extensions to the standard model,
in particular left-right symmetric models predicting a lower bound
on $\rm G_{M\overline{M}}$.

{\sc Acknowledgements. }
This work is supported in part by 
the German BMBF, 
the Swiss Nationalfond, 
the Russian FFR and a NATO research grant.
The collaboration is grateful to the staff 
at PSI for providing excellent working
conditions in a friendly atmosphere.

\begin{table}
Table 1: Muonium to antimuonium conversion is allowed in some
speculative extensions to
the standard model. The lower limit given
for minimal left-right 
symmetry corresponds to a muon neutrino mass limit of $m_{\nu \mu}
\leq$ 160 keV/c$^2$ \cite{assa94}.
\begin{tabular}{|p{10cm}|p{3.5cm}|p{1.5cm}|}
Model & limit & Ref.    \\ \hline \hline 
Minimal left-right symmetry with extended Higgs sector;
conversion through exchange of doubly charged Higgs boson $\Delta^{++}$ 
& $\rm G_{M\overline{M}}$ $\geq 2\cdot10^{-4}$ $\rm G_F$
& \cite{herc92} \\ \hline 
$Z_8$ model with forth generation of heavy 
leptons and radiative mass generation in leptonic sector;
conversion through exchange of neutral scaler boson $\Phi^{0}$
& $\rm G_{M\overline{M}}$ $\leq 10^{-2}$ $\rm G_F$ &
\cite{wong94} \\ \hline 
Supersymmetric model with broken R-parity;
conversion through exchange of $\tau$-sneutrino $\tilde{\nu_\tau}$
& $\rm G_{M\overline{M}}$ $\leq 10^{-2}$ $\rm G_F$ 
& \cite{moha92,halp93} \\ \hline
GUT models; conversion through exchange of doubly charged
dileptonic gauge boson $X^{\pm\pm}$ 
& $\rm G_{M\overline{M}}$ $\leq  0.13$ $\rm G_F$ & \cite{fuji94} \\
\end{tabular}
\end{table}

\begin{figure}[hbt]
   \centering\caption[]
        {\label{mmbarsetup}
        Top view of the new apparatus
	at PSI to search for
        muonium-antimuonium conversion. The observation of an energetic
	electron from the $\mu ^-$ decay
	in the antiatom in a magnetic spectrometer
	is required in coincidence with
	the detection of the positron,
	which is left behind
	from the atomic shell of the antiatom, on a MCP and at least one
	annihilation
	photons in a CsI calorimeter.
	}
\end{figure}

\begin{figure}[hbt]
        \centering \caption[] {
	\label{muonium}
	(a) The yield of muonium atoms in vacuum can be
	obtained from the distribution
	of time intervals between the incomming muon detected by 
	the beam counter and the observation of the decay of an atom
	in vacuum. The dashed line represents an exponentially decaying
	background.
	(b) An independent two-dimensional fit to the observed muon decays
	in the vacuum region downstream of the $SiO_2$ target
	as a function of the time and position gives the
	same muonium fraction. Logarithmic contours are
	displayed.
	}
\end{figure}

\begin{figure}[hbt]
   \centering\caption[]
        {\label{mbarresult}
         The number of events with identified energetic electron 
	 and positron as a function of
         (a) the distance of closest approach $R_{dca}$
	 between the electron track in the 
	 magnetic spectrometer and the back projection of the position 
	 measured at the MCP and (b) the difference of the
         positrons time of flight $t_{TOF}$ 
	 to the expected arrival time $t_{expected}$. 
	 The signal at earlier times corresponds 
	 to the allowed decay channel
	 $\mu \rightarrow  3e2\nu$ and Bhabha scattering.
	 It is smeared out due to different
	 acceleration voltages used. 
	 No event satisfied the
         required coincidence signature.
	 The dashed curves correspond to simulated
         signals scaled from the
	 relevant distributions recorded while monitoring 
	 the muonium production for ${\rm G_{M\overline{M}}}$=0.05 G$_F$.
         }
\end{figure}

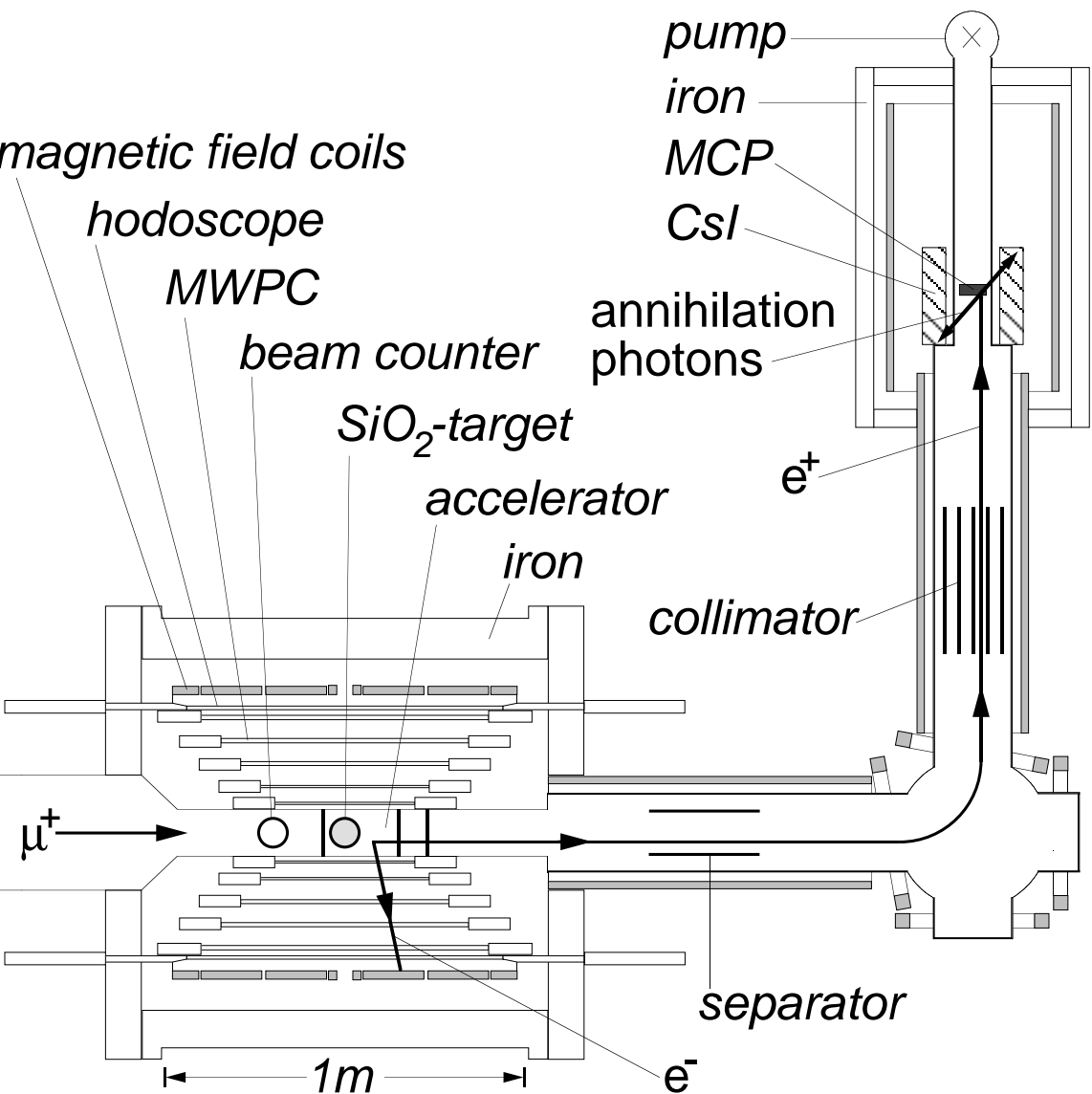
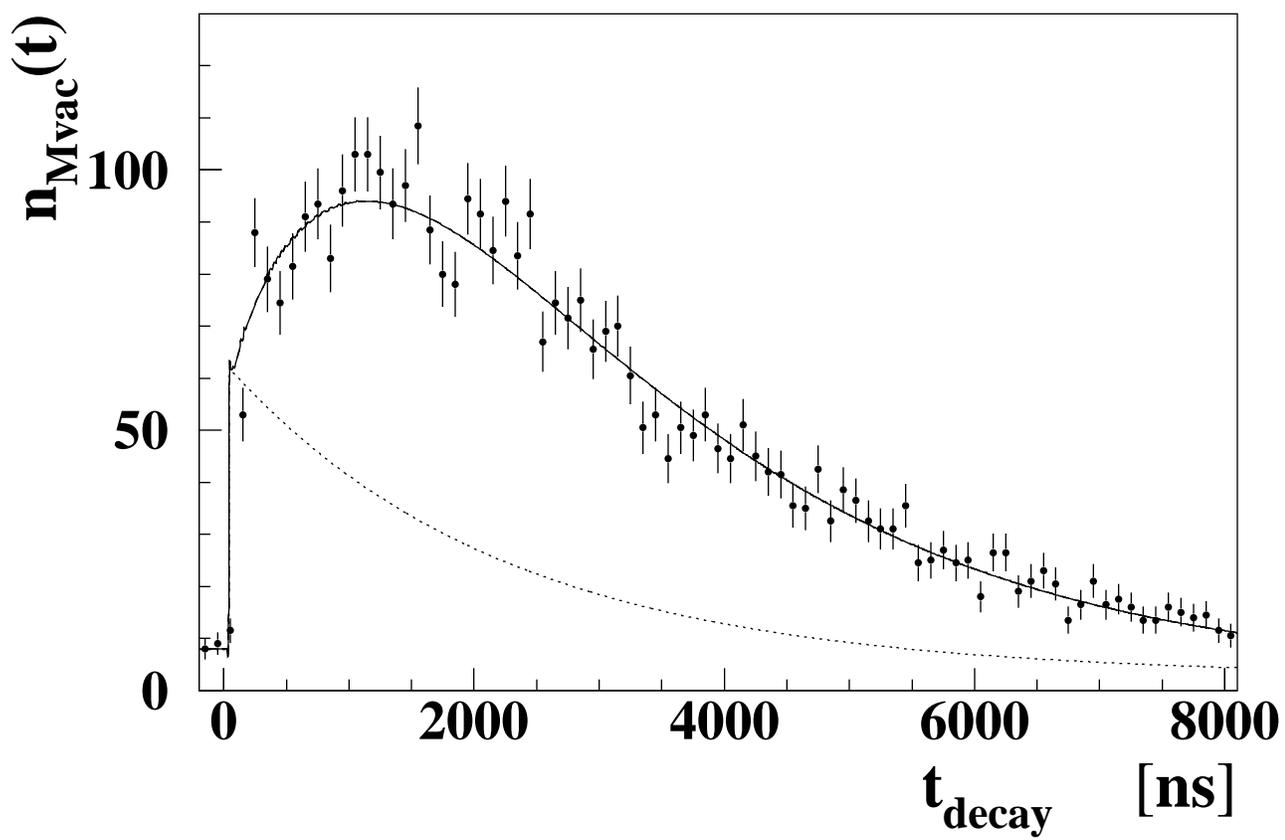
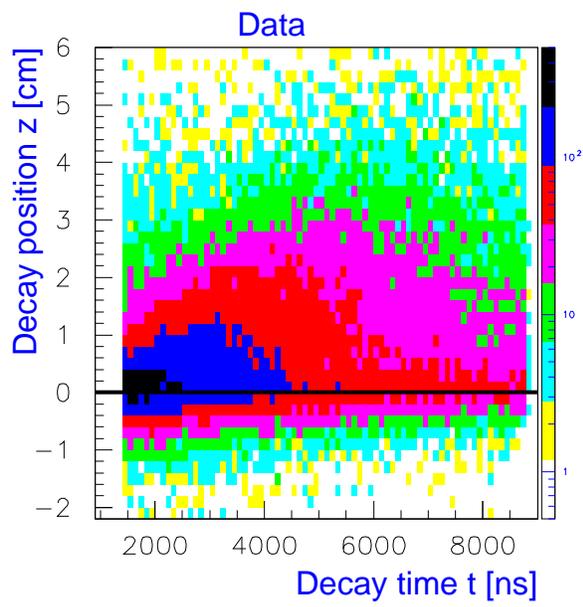
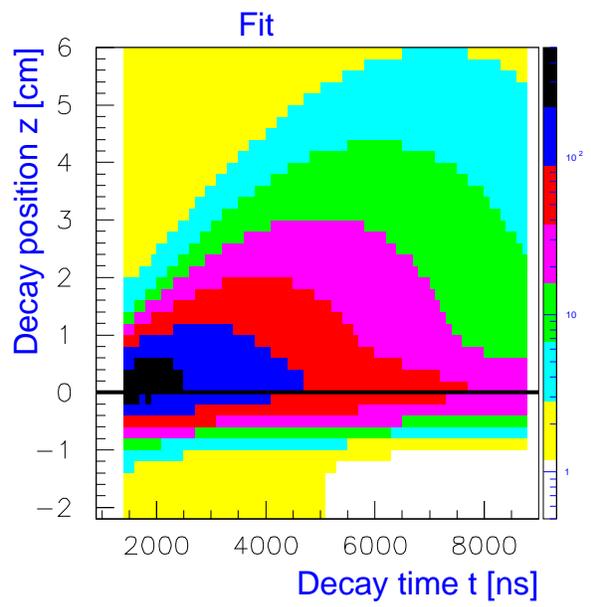
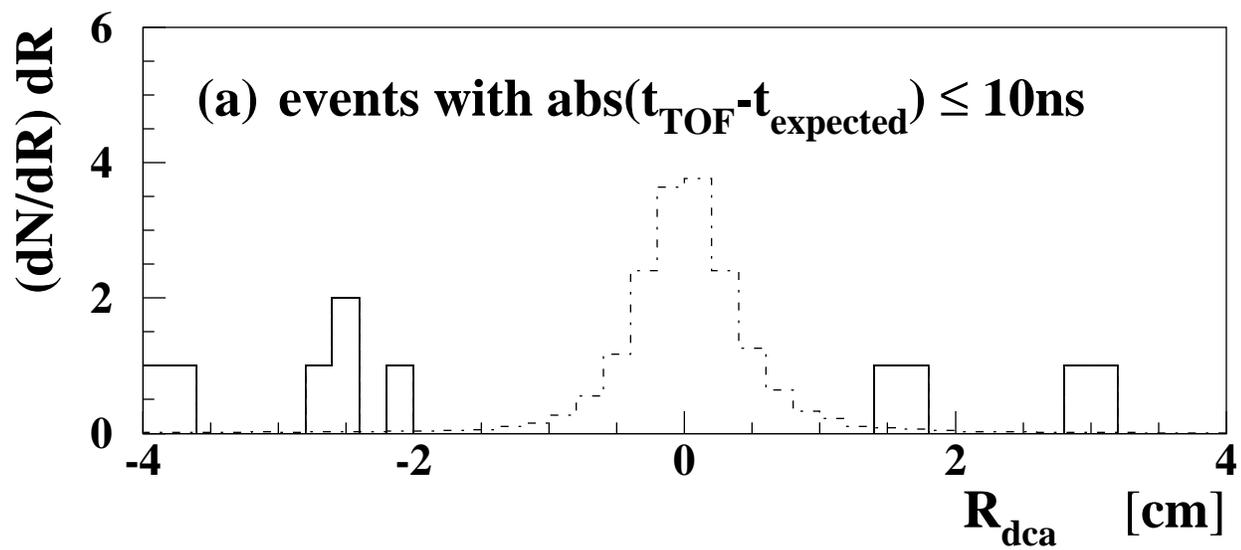
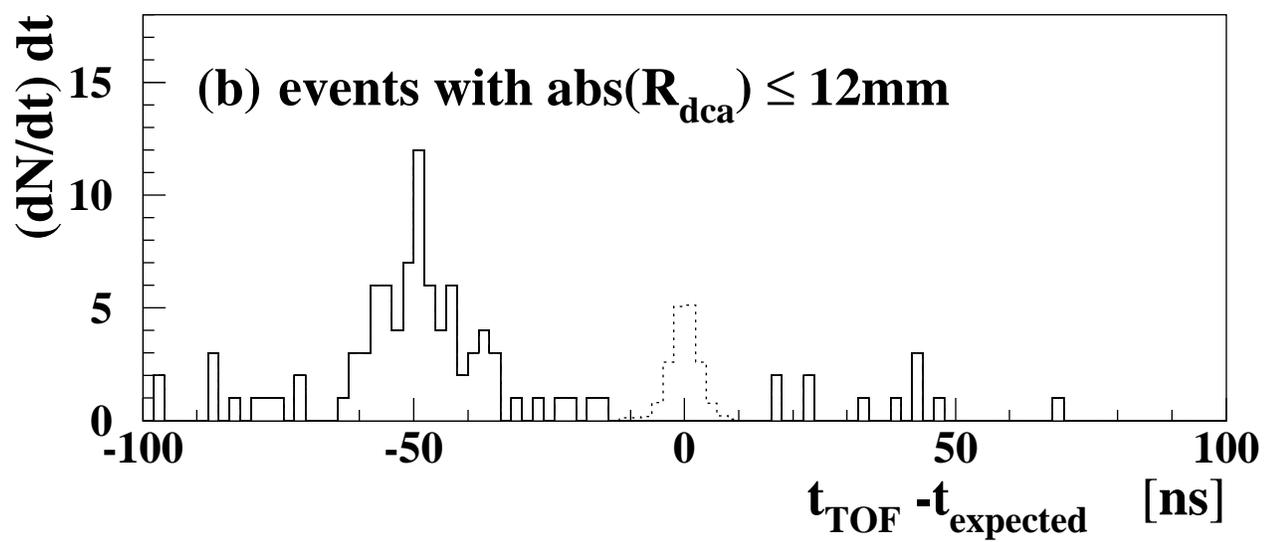
\end{document}